\documentclass[showpacs,preprintnumbers,twocolumn,
amsmath,amssymb,groupedaddress,superscriptaddress]{revtex4}
\usepackage{graphicx}
\usepackage{dcolumn}
\usepackage{bm}

\begin{document}

\preprint{}

\title{Scaling Functions and Superscaling in Medium and Heavy Nuclei}

\author{A.N. Antonov}
\affiliation{Institute for Nuclear Research and Nuclear Energy,
Bulgarian Academy of Sciences, Sofia 1784, Bulgaria}

\author{M.V. Ivanov}
\affiliation{Institute for Nuclear Research and Nuclear Energy,
Bulgarian Academy of Sciences, Sofia 1784, Bulgaria}

\author{M.K. Gaidarov}
\affiliation{Institute for Nuclear Research and Nuclear Energy,
Bulgarian Academy of Sciences, Sofia 1784, Bulgaria}

\author{E. Moya de Guerra}
\affiliation{Instituto de Estructura de la Materia, CSIC, Serrano
123, E-28006 Madrid, Spain}%
\affiliation{Departamento de Fisica Atomica, Molecular y Nuclear,
Facultad de Ciencias Fisicas, Universidad Complutense de Madrid,
E-28040 Madrid, Spain}

\author{P. Sarriguren}
\affiliation{Instituto de Estructura de la Materia, CSIC, Serrano
123, E-28006 Madrid, Spain}

\author{J.M. Udias}
\affiliation{Departamento de Fisica Atomica, Molecular y Nuclear,
Facultad de Ciencias Fisicas, Universidad Complutense de Madrid,
E-28040 Madrid, Spain}

\date{\today}

\begin{abstract}
The scaling function $f(\psi')$ for medium and heavy nuclei with
$Z\neq N$ for which the proton and neutron densities are not
similar is constructed within the coherent density fluctuation
model (CDFM) as a sum of the proton and neutron scaling functions.
The latter are calculated in the cases of $^{62}$Ni, $^{82}$Kr,
$^{118}$Sn, and $^{197}$Au nuclei on the basis of the
corresponding proton and neutron density distributions which are
obtained in deformed self-consistent mean-field Skyrme HF+BCS
method. The results are in a reasonable agreement with the
empirical data from the inclusive electron scattering from nuclei
showing superscaling for negative values of $\psi'$, including
those smaller than -1. This is an improvement over the
relativistic Fermi gas (RFG) model predictions where $f(\psi')$
becomes abruptly zero for $\psi'\leq -1$. It is also an
improvement over the CDFM calculations made in the past for nuclei
with $Z\neq N$ assuming that the neutron density is equal to the
proton one and using only the phenomenological charge density.
\end{abstract}


\maketitle

The studies of the scaling phenomenon which has been observed in
inclusive electron scattering from nuclei make it possible to gain
information about basic nuclear characteristics such as the local
density $\rho(r)$ and momentum distribution $n(k)$ in nuclei. This
concerns firstly the $y$-scaling
(e.g.~\cite{West75,Sick80,Pace82,Ciofi83,Day90}). It was found
also that a properly defined function $f(\psi')$ of another
scaling variable (the $\psi'$-variable) has a superscaling
behavior. The latter means that for $\psi'<0$ this function is
independent on the transfer momentum $q$ (at $q > 500$~MeV/c) and
on the mass number for a wide range of nuclei from $^4$He to
$^{197}$Au. This was firstly considered within the framework of
the RFG model
(e.g.~\cite{Alberico88,Donnelly99a,Donnelly99b,Maieron02,Barbaro04}).
As pointed out in~\cite{Donnelly99b}, however, the actual nuclear
dynamical content of the superscaling is more complex than that
provided by the RFG model. It was observed that the experimental
data have a superscaling behavior for large negative values of
$\psi'$ (up to $\psi'\approx-2$), while the predictions of the RFG
model are for $f(\psi') = 0$ at  $\psi'\leq-1$. This imposes the
consideration of the superscaling in realistic finite systems.
Such works were performed~\cite{Antonov04,Antonov05} in the
CDFM~\cite{Antonov88,Antonov79,Antonov89}, which is related to the
$\delta$-function limit of the generator-coordinate
method~\cite{Antonov04,Griffin57}. The calculated CDFM scaling
function $f(\psi')$ agrees with the available experimental data
from the inclusive electron scattering for $^4$He, $^{12}$C,
$^{27}$Al, $^{56}$Fe and, approximately, for $^{197}$Au for
various values of the transfer momentum $q = 500$, 1000,
1650~MeV/c~\cite{Antonov04} and 1560~MeV/c~\cite{Antonov05},
showing superscaling for negative values of $\psi'$ including also
those smaller than $-1$ (in contrast to the RFG model result). It
was shown in~\cite{Antonov04,Antonov05} that the superscaling in
nuclei can be explained quantitatively on the basis of the similar
behavior of the high-momentum components of the nucleon momentum
distributions in light, medium and heavy nuclei. It is known that
the latter is related to the effects of the short-range and tensor
nucleon-nucleon correlations in nuclei (see,
e.g.~\cite{Antonov88}). Our scaling function was obtained starting
from that in the RFG
model~\cite{Alberico88,Donnelly99a,Donnelly99b} in two equivalent
ways, on the basis of the local density distribution and of the
nucleon momentum distribution. This gives a good opportunity to
study simultaneously the role of the nucleon-nucleon correlations
included in $\rho(r)$ and $n(k)$ in the case of the superscaling
phenomenon.

Here we would like to emphasize, however, that
in~\cite{Antonov04,Antonov05} we encountered some difficulties to
describe within the CDFM the superscaling in the case of $^{197}$Au
which was the most heavy nucleus considered. We related this to the
particular $A$-dependence of $n(k)$ in the model that does not lead
to realistic high-momentum components of the momentum distribution
in the heaviest nuclei. We followed in~\cite{Antonov04,Antonov05} a
somewhat artificial way to ``improve'' the high-momentum tail of
$n(k)$ in $^{197}$Au by taking the value of the diffuseness
parameter $b$ in the Fermi-type charge density distribution of this
nucleus to be $b =1$~fm instead of the value $b = 0.449$~fm (as
obtained from electron elastic scattering experiments,
see~\cite{Patterson03}). In such a case the high-momentum tail of
$n(k)$ for $^{197}$Au in CDFM becomes similar to those of $^{4}$He,
$^{12}$C, $^{27}$Al, and $^{56}$Fe nuclei and this leads to a good
agreement of the scaling function $f(\psi')$ with the data also for
$^{197}$Au. Discussing this in~\cite{Antonov04} we pointed out,
however, that all the nucleons may contribute to $f(\psi')$ for the
transverse electron scattering and this could reflect on the
diffuseness of the matter density for a nucleus like $^{197}$Au
whose value can be different from that of the charge density used in
our previous works~\cite{Antonov04,Antonov05}.

The aim of the present work is to apply the CDFM by using both
proton and neutron densities for medium and heavy nuclei (for which
$Z \neq N$) in contrast to our previous approach, in which we
assumed that the neutron density was equal to that of protons and we
used only the phenomenological charge density~\cite{Patterson03}. In
our work now the total scaling function $f(\psi')$ will be a sum of
two scaling functions, those for protons and neutrons.

In~\cite{Antonov05} the CDFM scaling function $f(\psi')$ was given
in two equivalent ways, firstly, by means of the density
distribution
\begin{equation}\label{eq1-antonov}
f(\psi')= \int\limits_{0}^{\alpha/(k_{F}|\psi'|)}dR |F(R)|^{2}
f_{RFG}(R,\psi'),
\end{equation}
where
\begin{equation}\label{eq2-antonov}
|F(R)|^{2}=-\frac{1}{\rho_{0}(R)} \left. \frac{d\rho(r)}{dr}\right
|_{r=R},
\end{equation}
\begin{equation}\label{eq3-antonov}
\rho_{0}(R)=\frac{3A}{4\pi R^{3}},\quad \alpha=\left(\frac{9\pi
A}{8}\right )^{1/3}\!\!\simeq 1.52A^{1/3},
\end{equation}
\begin{align}\label{eq4-antonov}
&f_{RFG}(R,\psi') =  \frac{3}{4} \Bigg[ 1-\Big(
\frac{k_FR|\psi'|}{\alpha} \Big)^{2}\Bigg] \Bigg\{ 1+ \Big(
\frac{Rm_N}{\alpha}\Big)^2\nonumber\\
&\times \Big( \frac{k_FR|\psi'|}{\alpha} \Big)^2\! \Bigg[2\!+\!
\Big( \frac{\alpha}{Rm_N} \Big)^2\!\!\!-\! 2\sqrt{1+ \Big(
\frac{\alpha}{Rm_N} \Big)^2}\Bigg]\! \Bigg\},
\end{align}
($m_N$ being the nucleon mass), and secondly, by means of the
momentum distribution
\begin{equation}\label{eq5-antonov}
f(\psi')= \int\limits_{k_{F}|\psi'|}^{\infty} d\overline{k}_F
|G(\overline{k}_F)|^2 f_{RFG}(\overline{k}_F, \psi'),
\end{equation}
where
\begin{equation}\label{eq6-antonov}
|G(\overline{k}_F)|^2=- \frac{1}{n_0(\overline{k}_F)}\left.
\frac{dn(p)}{dp}\right |_{p=\overline{k}_F}
\end{equation}
and
\begin{equation}\label{eq7-antonov}
n_0(\overline{k}_F)= \frac{3A}{4\pi {\overline{k}_F}^3} .
\end{equation}
In Eq.~(\ref{eq5-antonov}) the RFG scaling function
$f_{RFG}(\overline{k}_F,\psi')$ can be obtained from
$f_{RFG}(R,\psi')$ (Eq.~(\ref{eq4-antonov})) by changing there
$\alpha/R$ by $\overline{k}_F$. In Eqs.~(\ref{eq1-antonov}),
(\ref{eq4-antonov}), and (\ref{eq5-antonov}) the Fermi momentum
$k_F$ is not a free fitting parameter for different nuclei as in the
RFG model, but it is calculated in the CDFM for each nucleus using
the corresponding expressions:
\begin{multline}\label{eq8-antonov}
k_F= \int\limits_{0}^{\infty} dR k_{F}(R)|F(R)|^2=
\int\limits_{0}^{\infty} dR \frac{\alpha}{R}|F(R)|^{2}=\\=
\frac{4\pi(9\pi)^{1/3}}{3A^{2/3}} \int\limits_{0}^{\infty} dR
\rho(R) R
\end{multline}
when
\begin{equation}\label{eq9-antonov}
\lim_{R\rightarrow \infty} \big[ \rho(R)R^2 \big]=0
\end{equation}
is fulfilled and
\begin{equation}\label{eq10-antonov}
k_F= \frac{16\pi}{3A} \int\limits_0^\infty d\overline{k}_F
n(\overline{k}_F ) {\overline{k}_F}^3
\end{equation}
when
\begin{equation}\label{eq11-antonov}
\lim_{\overline{k}_F\rightarrow \infty}\big[
n(\overline{k}_F){\overline{k}_F}^4 \big]=0
\end{equation}
is fulfilled.

In~\cite{Antonov04,Antonov05} we used the charge density
distributions to determine the weight function $|F(R)|^2$ in
calculations of $f(\psi')$ from
Eqs.~(\ref{eq1-antonov})--(\ref{eq4-antonov}) and
(\ref{eq8-antonov}). In the present work we assume that the reason
why the CDFM does not work properly in the case of $^{197}$Au is
that we use in~\cite{Antonov04,Antonov05} only the charge density,
while this nucleus has many more neutrons than protons ($N=118$ and
$Z= 79$), and therefore proton and neutron densities may differ
considerably. In this case the proton $f_p(\psi')$ and neutron
$f_n(\psi')$ scaling functions will be given by the contributions of
the proton and neutron densities $\rho_p(r)$ and $\rho_n(r)$,
correspondingly:
\begin{equation}\label{eq12-antonov}
f_{p(n)}(\psi^{\prime})=\!\!\!\!\int\limits_{0}^{\alpha_{p(n)}/(k^{p(n)}_{F}|\psi^{\prime}|)}\!\!\!\!dR
|F_{p(n)}(R)|^{2}f_{RFG}^{p(n)}(R,\psi'),
\end{equation}
where the proton and neutron weight functions are obtained from the
corresponding proton and neutron densities
\begin{equation}\label{eq13-antonov}
\left|F_{p(n)}(R)\right|^2=-\dfrac{4\pi
R^3}{3Z(N)}\left.\dfrac{d\rho_{p(n)}(r)}{dr}\right|_{r=R},
\end{equation}
\begin{equation}\label{eq14-antonov}
\alpha_{p(n)}=\left(\dfrac{9\pi Z(N)}{4}\right)^{1/3},
\end{equation}
\begin{equation}\label{eq15-antonov}
\int\limits_{0}^{\infty}\rho_{p(n)}(\vec{r})d\vec{r}=Z(N),
\end{equation}
and the Fermi momentum for the protons and neutrons is given by
\begin{equation}\label{eq16-antonov}
k_{F}^{p(n)}=\alpha_{p(n)}\int\limits_{0}^{\infty}dR
\frac{1}{R}|{F}_{p(n)}(R)|^{2}.
\end{equation}
The RFG proton and neutron scaling functions
$f_{RFG}^{p(n)}(R,\psi')$ have the form of Eq.~(\ref{eq4-antonov}),
where $\alpha$ and $k_F$ are changed by $\alpha_{p(n)}$ from
Eq.~(\ref{eq14-antonov}) and $k_F^{p(n)}$ from
Eq.~(\ref{eq16-antonov}), correspondingly. The normalizations of the
functions are as follows:
\begin{gather}
\int\limits_{0}^{\infty}|F_{p(n)}(R)|^{2}dR=1,\label{eq17-antonov}\\
\int\limits_{-\infty}^{\infty}f_{p(n)}(\psi^{\prime})d\psi^{\prime}=1.\label{eq18-antonov}
\end{gather}
Then the total scaling function can be expressed by means of both
proton and neutron scaling functions:
\begin{equation}\label{eq19-antonov}
f(\psi^{\prime})=\dfrac{1}{A}\big(Zf_p(\psi^{\prime})+Nf_n(\psi^{\prime})\big)
\end{equation}
and is normalized to unity.

The same consideration can be performed equivalently on the basis of
the nucleon momentum distributions for protons $n^p(k)$ and $n^n(k)$
presenting $f(\psi')$ by the sum of proton and neutron scaling
functions (\ref{eq19-antonov}) calculated similarly to
Eqs.~(\ref{eq12-antonov})--(\ref{eq19-antonov}) (and to
Eqs.~(\ref{eq5-antonov}), (\ref{eq6-antonov}), (\ref{eq10-antonov}),
and (\ref{eq11-antonov})):
\begin{equation}\label{eq20-antonov}
f_{p(n)}(\psi')=\!\! \int\limits_{k^{p(n)}_{F}|\psi'|}^{\infty}\!\!
d\overline{k}_F |G_{p(n)}(\overline{k}_F)|^2
f^{p(n)}_{RFG}(\overline{k}_F, \psi'),
\end{equation}
where
\begin{equation}\label{eq21-antonov}
|G_{p(n)}(\overline{k}_F)|^2=- \frac{4\pi
{\overline{k}_F}^3}{3Z(N)}\left. \frac{dn^{p(n)}(p)}{dp}\right
|_{p=\overline{k}_F}
\end{equation}
with $f_{RFG}^{p(n)}({\overline{k}_F},\psi')$ containing
$\alpha_{p(n)}$ from Eq.~(\ref{eq14-antonov}) and $k_F^{p(n)}$
calculated by
\begin{equation}\label{eq22-antonov}
k^{p(n)}_F=  \int\limits_0^\infty d\overline{k}_F \overline{k}_F
|G_{p(n)}(\overline{k}_F)|^2.
\end{equation}

We calculate the scaling function for several examples, for the
medium stable nuclei $^{62}$Ni and $^{82}$Kr and for the heavy
nuclei $^{118}$Sn and $^{197}$Au following
Eqs.~(\ref{eq12-antonov})--(\ref{eq19-antonov}) using the
corresponding proton and neutron densities obtained in deformed
self-consistent mean-field (HF+BCS) calculations with
density-dependent Skyrme effective interaction (SG2) using a large
harmonic-oscillator basis with 11 major
shells~\cite{Sarriguren99,Vautherin73}.

\begin{figure}[t]
\includegraphics[width=80mm]{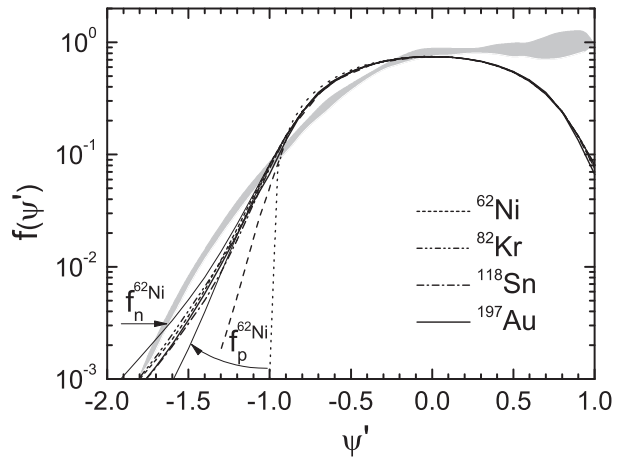}
\caption{\label{fig1-antonov} Scaling function $f(\psi')$ calculated
in the CDFM for $^{62}$Ni, $^{82}$Kr, $^{118}$Sn, and $^{197}$Au.
The results are obtained using
Eqs.~(\ref{eq12-antonov})--(\ref{eq19-antonov}). The experimental
data for $^{4}$He, $^{12}$C, $^{27}$Al, $^{56}$Fe, and $^{197}$Au at
$q=1000$~MeV/c taken from~\cite{Donnelly99b} are shown by the shaded
area. The RFG result is presented by dotted line. The results of the
calculations for $^{197}$Au~\cite{Antonov04} by means of
Eqs.~(\ref{eq1-antonov})--(\ref{eq4-antonov}) using the Fermi-type
charge density~\cite{Patterson03} are shown by dashed line.}
\end{figure}

The results of the calculations of $f(\psi')$ for $^{62}$Ni,
$^{82}$Kr, $^{118}$Sn, and $^{197}$Au for $q=1000$~MeV/c are given
in Fig.~\ref{fig1-antonov} and are compared  with the experimental
data (presented by a gray area and taken from~\cite{Donnelly99b})
obtained for $^{4}$He, $^{12}$C, $^{27}$Al, $^{56}$Fe, and
$^{197}$Au. The scaling functions are in a reasonable agreement
with the data, which was not the case for $^{197}$Au calculated
in~\cite{Antonov04} by using the experimental Fermi-type charge
density only with parameter values $b=0.449$~fm and $R=6.419$~fm
from~\cite{Patterson03} (see also Fig.~2 of
Ref.~\cite{Antonov04}). It must be stressed that the theoretical
proton densities used in the present calculations are close to the
experimental charge densities. At the same time we note also the
improvement in comparison with the RFG model result in which
$f(\psi')=0$ for $\psi'\leq-1$. As an example the proton and
neutron scaling functions for $^{62}$Ni are also given in
Fig.~\ref{fig1-antonov}, which clearly illustrates the different
tails of the proton and neutron scaling functions when $Z\neq N$,
as well as their role in building up the observed scaling
function.

In conclusion, we point out that the scaling function $f(\psi')$
for nuclei with $Z\neq N$ for which the proton and neutron
densities are not similar has to be expressed by the sum of the
proton and neutron scaling functions. The latter can be calculated
within the CDFM on the basis of the knowledge (obtained
theoretically and/or experimentally) of the corresponding proton
and neutron local density distributions or momentum distributions.
We should also point out that the agreement with experiment is
quite reasonable given that no adjustable parameter at all has
been used in the present calculations.

\begin{acknowledgments}
Three of the authors (A.N.A., M.V.I. and M.K.G.) are thankful to the
Bulgarian National Science Foundation for partial support under
Contracts No.$\Phi$-1416 and $\Phi$-1501. This work was partly
supported by funds provided by DGI of MCyT (Spain) under Contract
Nos. FIS 2005-00640, BFM 2000-0600, and BFM-04147-C02-01 and by the
Agreement (2004 BG2004) between the CSIC (Spain) and the Bulgarian
Academy of Sciences. One of us (E.M.G.) is indebted to Prof.
T.W.~Donnelly for valuable comments and suggestions.
\end{acknowledgments}

\end{document}